\documentclass[onecolumn,showpacs,superscriptaddress,amsmath,amssymb,preprint,aps]{revtex4}
\usepackage{latexsym}
\usepackage[mathscr]{eucal}
\usepackage{amsthm,amsxtra,amscd,upref}
\usepackage{layout,bm,dcolumn}
\usepackage{graphicx,color}
\usepackage{calc}
\newcommand{\DST}[1]{{\ensuremath{\displaystyle{#1}}}}
\newcommand{\DSF}[2]{{\ensuremath{\displaystyle{\frac{\DST{#1}}{\DST{#2}}}}}}
\newcommand{\vk}[1]{{\ensuremath{{\bm{#1}}}}}

\newcommand{\lrb}[1]{{\ensuremath{\left({#1}\right)}}}
\newcommand{\lrs}[1]{{\ensuremath{\left[{#1}\right]}}}

\newcommand{\lrv}[1]{{\ensuremath{\left|{#1}\right|}}}
\newcommand{\lra}[1]{{\ensuremath{\left\langle{#1}\right\rangle}}}

\newcommand{\vrr}{{\ensuremath{\vk{r}}}}

\newcommand{\vqr}{{\ensuremath{\vk{q}}}}

\newcommand{\IMA}{{\ensuremath{\mathrm{i}}}}
\newcommand{\EE}{{\ensuremath{\mathrm{e}}}}

\newcommand{\DDD}[2]{{\ensuremath{\mathrm{d}^{#1}{#2}\, }}}

\newcommand{\haf}{{\ensuremath{\scriptstyle{\frac{1}{2}}}}}

\newcommand{\eref}[1]{Eq.~(\ref{#1})}

\newcommand{\eeref}[2]{Eqs.~(\ref{#1},\ref{#2})}

\newcommand{\sref}[1]{Sec.~\ref{#1}}
\newcommand{\cref}[1]{Chap.~\ref{#1}}

\newcommand{\degC}{{\ensuremath{{}^{\mathrm{o}}}}}

\begin{document}
\title{ 
Cube-shape diffuse scattering and the ground state of $\mathrm{BaMg}_{1/3}\mathrm{Ta}_{2/3} \mathrm{O}_3$ 
}

\author{A.~Cervellino}
\email{antonio.cervellino@psi.ch} 
\affiliation{Swiss Light Source, Paul Scherrer Institut, CH-5232 Villigen, Switzerland}
\affiliation{Laboratory for Neutron Scattering, PSI, CH-5232 Villigen, Switzerland}
\author{S.N.~Gvasaliya}
\affiliation{Laboratory for Neutron Scattering, PSI, CH-5232 Villigen, Switzerland}
\affiliation{Neutron Scattering and Magnetism Group, Institute for Solid State Physics, ETH Z\"urich, Z\"urich, Switzerland}
\author{B.~Roessli}
\affiliation{Laboratory for Neutron Scattering, PSI, CH-5232 Villigen, Switzerland}
\author{G.M.~Rotaru}
\affiliation{Laboratory for Neutron Scattering, PSI, CH-5232 Villigen, Switzerland}
\author{R.A.~Cowley}
\affiliation{Clarendon Laboratory, Department of Physics, Oxford University, Parks Road, Oxford, UK}
\author{S.G.~Lushnikov}
\affiliation{Ioffe Physical Technical Institute, 26 Politekhnicheskaya, 194021 St. Petersburg, Russia}
\author{T.A.~Shaplygina}
\affiliation{Ioffe Physical Technical Institute, 26 Politekhnicheskaya, 194021 St. Petersburg, Russia}
\author{A. Bossak}
\affiliation{European Synchrotron Radiation Facility, Bo{\^\i}te Postal 220, F-38043 Grenoble Cedex, France}
\author{D. Chernyshov}
\affiliation{Swiss-Norwegian Beam Lines at ESRF, 38043 Grenoble, France}

\date{\today}
\begin{abstract}
A quite unusual diffuse scattering phenomenology was observed in the single-crystal X-ray diffraction 
pattern of cubic perovskite BMT ($\mathrm{BaMg}_{1/3}\mathrm{Ta}_{2/3}\mathrm{O}_3$). 
The intensity of the scattering is parametrized as a set of cube-like objects located at the centers of reciprocal space 
unit cells, resembling very broad and cubic-shaped (1/2,1/2,1/2)-satellites. 
BMT belongs to perovskites of formula AB$'_{1/3}$B$''_{2/3}$O$_{3}$ (A=Mg, B$'=$Ta, B$''=$Mg).  
The cubes of the intensity can be attributed to the partial correlations of the occupancies of the B site. 
The pair correlation function is the Fourier transform of the diffuse 
scattering intensity and the latter's idealized form yields the unusual property of a power-law correlation decay with distance.  
Up to now this is observed only in a few exotic instances of magnetic order or nematic crystals. Therefore it cannot 
be classified as a short-range order phenomenon, 
as in most situations originating diffuse scattering. A Monte-Carlo search in configuration space 
yielded solutions that reproduce faithfully the observed diffuse scattering. 
Analysis of the results in terms of the electrostatic energy and the entropy point to this phase of BMT 
as a metastable state, kinetically locked, which could be the equilibrium state 
just below the melting point. 
\end{abstract}
 
\maketitle

\section{Introduction.}

An experiment using X-ray single crystal diffraction at the Swiss-Norwegian Beam Lines (SNBL) of 
European Synchrotron Radiation Facility (ESRF) on a single crystal of cubic perovskite BMT 
($\mathrm{BaMg}_{1/3}\mathrm{Ta}_{2/3}\mathrm{O}_3$) 
has shown an interesting diffuse scattering pattern, consisting of cube-shaped 
domains of nearly constant intensity, with the centers located on the (1/2,1/2,1/2) points of the 
reciprocal lattice and with an edge of $\sim 1/3$ of a reciprocal lattice unit (rlu). All 
the cubes have the same intensity level, apart from the usual slow modulations 
due to atomic and thermal factors. 

BMT belongs to perovskites of 
formula AB$'_{1/3}$B$''_{2/3}$O$_{3}$ (A=Ba, B$'=$Ta, B$''=$Mg)
where different species on the B-site can be ordered to various degrees. The scattering 
depends on the site occupancy and when the ordering is directly related to the pair 
correlation function of the different atomic sites~\cite{Krivoglaz1}. 
In fact, BMT in its cubic phase is necessarily intrinsically disordered in the B-site. In 
this paper we shall mainly explore the chemical order at the B-site and show that the cube-like 
diffuse scattering can be fully explained in this way. 

Lattice deformations and displacements, 
due to different atomic radii of Mg and Ta (see Ref.~\cite{Lufaso04}), are expected. 
So there is additional diffuse scattering originating from these correlated displacements. 
In fact streak-like diffuse scattering - relatively weaker - has been found at the foot of the cubic Bragg peaks, see Fig.~\ref{fig1}. 
We shall address also this point in a future paper. 

The simplest possibility for B-site ordering is the formation of subregions enriched in either Mg or Ta cations. 
This possibility, however, has been thoroughly ruled out as energetically unfavourable in  AB$'_{1/3}$B$''_{2/3}$O$_{3}$ 
perovskites~\cite{Vieh95,Yan98} because of the large charges in the subregions. There are then different homogeneous ways 
of ordering the B$'$-B$''$ cations. 
These are:
\begin{itemize}
\item[i)]{homogeneous random distribution of B-cations;}
\item[ii)]{correlated distribution with occupation numbers being a well-defined function of distance between two 
lattice nodes, but without periodicity;}
\item[iii)]{periodic long range ordering of the Mg/Ta concentrations in different sublattices.}
\end{itemize}
Situation i) would give an almost wavevector-independent diffuse scattering, but this is not observed.  
B-site ordering of type iii) has been observed in BMT, with simultaneous lowering of the cubic symmetry. In fact, 
BMT may exist in an ordered trigonal structure ($P\overline{3}m1$) modification~\cite{Galasso63,Kawa83,AkDa98,Lufaso04,cervellino_I} and it 
has been shown that this modification~\cite{Lufaso04} mainly entails perfect periodic ordering of the B-site. The trigonal lattice 
is essentially a threefold superlattice of the cubic one, with extremely small geometric distortions, realized by 
stacking periodically a sequence of one Mg plane and two Ta planes orthogonally to one of the cubic $\lra{1,1,1}$ directions. 
The atomic sites are also very close to those of the underlying cubic sublattice, with only small atomic displacements 
consistent with the reduced symmetry. It is noteworthy that the trigonal BMT superstructure - 
usually cast unto four equally populated domain orientations with the threefold axis along either of the cubic 
$\langle 111\rangle$ directions, so that the average symmetry is conserved - 
can be considered as a commensurate distortion of cubic BMT. In this framework, the modulation contributes 
eight additional ($\pm1/3, \pm1/3, \pm1/3$) satellite Bragg peaks to the diffraction pattern, and these coincide with 
the vertices of the diffuse scattering cubes. 
Another partial B-site ordering type of class iii) has been observed~\cite{Vieh95,Yan98} in 
other AB$'_{1/3}$B$''_{2/3}$O$_{3}$ cubic perovskites (PMT, with A=Pb, B$'=$Ta, B$''=$Mg). 
These have (1/2,1/2,1/2) satellites and correspond to a partial B-site ordering with a 2$\times$2$\times$2 $fcc$ 
superstructure unit cell. The B-sites are of two types, one of pure B$''$ and the other containing the rest 
of B$''$ and all of B$'$ cations, and the two B-site types 
arrange in a NaCl-type structure that consists of two inter-penetrating fcc lattices. 
These satellites coincide now with the center of the diffuse scattering cubes observed for BMT. 

Diffuse scattering cubes straddle the reciprocal space location of the (1/2,1/2,1/2) satellites 
and the ($\pm1/3, \pm1/3, \pm1/3$) satellites identifying the two kinds of periodically ordered type iii)  
B-site structures. Although this diffuse scattering is not Bragg-like, in a sense that will be specified 
in~\sref{sec:smooth}, this suggests that the diffuse cubes result from B-site ordering as well, 
possibly from a non-periodic order of type ii). 

We stress here that the observed cubes of diffuse intensity are not ideal, as careful inspection (\emph{cf.} Fig.~\ref{fig1}) 
shows that "cubes" have a minimum in the center about 10\%{} lower than at the periphery 
and the corresponding intensity distribution is not a flat-top. We see in particular that intensity tends to increase 
near the vertices (the ($\pm1/3, \pm1/3, \pm1/3$) satellites of the trigonal superstructure) and decrease in the center - 
the (1/2,1/2,1/2) satellite of the hypothetical partly ordered NaCl-type superstructure, indicating an embryo of transition 
to the trigonal phase. We must therefore consider the phase with idealized cubes 
as a limiting situation that may not be perfectly realized in our sample. 
As we argue in the following that the phase with ideal diffuse cubes 
is not the ground state at room temperature but only a metastable phase, kinetically hindered from transforming to 
the trigonal ground state, this is perfectly possible, as a germ of transition may have occurred due to finite quenching speed. 
However, in this framework, the limiting phase with ideal diffuse cubes is a key to understanding the B-site ordering phenomena 
in BMT, and therefore we will deal mainly with this phase.

\subsection{A few remarks}

As we are discussing the multiple phases of BMT, 
we should comment on an important point about phase stability. 
In the thermodynamic sense, a stable structure observed in an experiment corresponds to a 
minimum of the free energy at the 
conditions of crystal formation. At ambient conditions the Mg-Ta distribution is quenched and 
experimentally observed configurations may not necessarily correspond to a true ground state. 
A disordered or incompletely ordered state can be stabilized at high 
temperature by a strong entropy term and could be a metastable state at lower T. 
In fact, at room temperature the diffusion of Mg and Ta is practically nonexistent, 
so that the non-ergodic effects in the scattering experiments can be neglected. 
 
If the first Born approximation is valid and the scattering can be considered as elastic, 
the scattered intensity is the Fourier transform of the scattering density's pair (two-body) 
correlation function~\cite{Guinier,WarrenX,Squires}. The scattering density is the atomic structure, 
and for X-rays it is the sum of the single atom electron densities and the only degrees of freedom 
come from the spatial distribution (structure) of the atoms themselves. 
The pair correlation function (crystallographers name it the "Patterson function") is the information 
contained in the scattering pattern. 
A natural ambiguity arises in the relationship between the atomic structure and the pair correlation function, as 
there may be many atomic structures that have the same pair correlation function~\cite{Patterson44,MerminIco}. 
In some cases, this is not an accidental degeneracy. When the internal energy $U$ depends only on the pairwise 
interactions, different structures with identical pair correlation functions are isoenergetic. At any finite 
temperature $T$, the multiplicity of configurations increases the entropy $S$ and this in turn decreases 
the free energy, $G=U-TS$, to the 
advantage of the stability of the (statistically defined) structure. We intend to show that 
the B-site order in BMT, that gives rise to the cubes of the diffuse intensity belongs to this case, 
based on the fact that pair (electrostatic) interactions between the B cations are sufficient to explain 
the phase diagram of the different BMT phases which derive from the ordering of B-sites. 

The pair (two-body) and the set of the higher order (many-body) correlation functions can 
fully describe any structure in an unambiguous way, including stochastically 
defined ones~\cite{Krivoglaz1,MerminIco}. Considering 
higher order correlation functions is necessary, however, only when 
they make important contributions to the internal energy. 
This is sometimes the case, for instance, for structures that show 
displacive diffuse scattering~\cite{Cowley68,Dietrich89a,Dietrich89b}. 
In any case, higher correlation functions are also a way of \emph{describing} a structure 
in the crystal-chemical sense. This is especially convenient when the structure itself cannot be 
easily represented because it is only statistically defined. However, simpler ways of presenting the 
crystal-chemical informations are also possible, see~\sref{MCsol}.

Our approach is as follows. The pair correlation function that is related to the diffuse cubes can 
be simply analyzed in terms of the B-site chemical order as a site concentration pair correlation 
function~\cite{Krivoglaz1}. This turns out to have a simple analytical form that shows a rather peculiar 
power-law decrease with distance that is observed in only a few cases for rather exotic systems. 

Monte-Carlo simulations were performed for a large scale in order to find the B-site configurations 
that give the same pair correlation as the diffuse intensity. Successfully determined configurations  
were then analyzed in terms of a long-range order parameter showing that these partially ordered configurations 
that are compatible with the observed scattering phenomena are not distinguishable from completely disordered systems. 
Next, we consider that BMT is an ionic compound and that the electrostatic energy is the main term in 
the crystal Hamiltonian. The electrostatic energy of a rigid system depends only on pair interactions and 
the electrostatic energies and the entropy terms for the different BMT phases were calculated. These gave a 
good correspondence with the observed phase diagram, confirming that 
the cubic phase must be a high-temperature ground state of BMT that is metastable at room temperature.

\section{Experimental}

The single crystals of BMT were synthesized according to the procedure described by 
Galasso and Pinto~\cite{gallaso1965}. The powder of BMT was prepared according to 
two-stage synthesis described in Ref.~\cite{gvasaliya2003} and is essentially the same as used in the 
neutron diffraction study~\cite{Gvas04_2}. 

\begin{figure}[!hbt]
\begin{center}
{\includegraphics[height=0.49\textwidth]{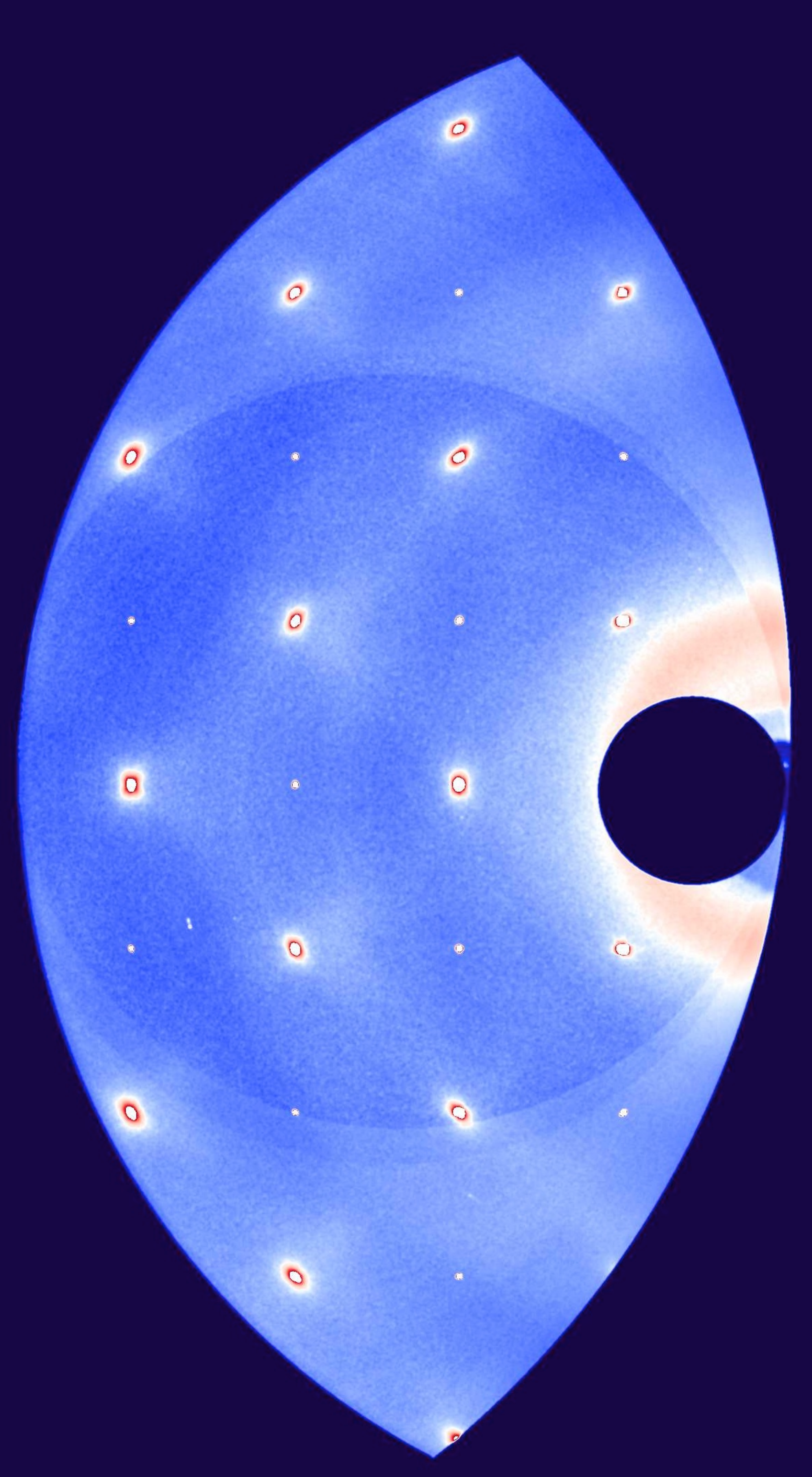}}
{\includegraphics[height=0.49\textwidth]{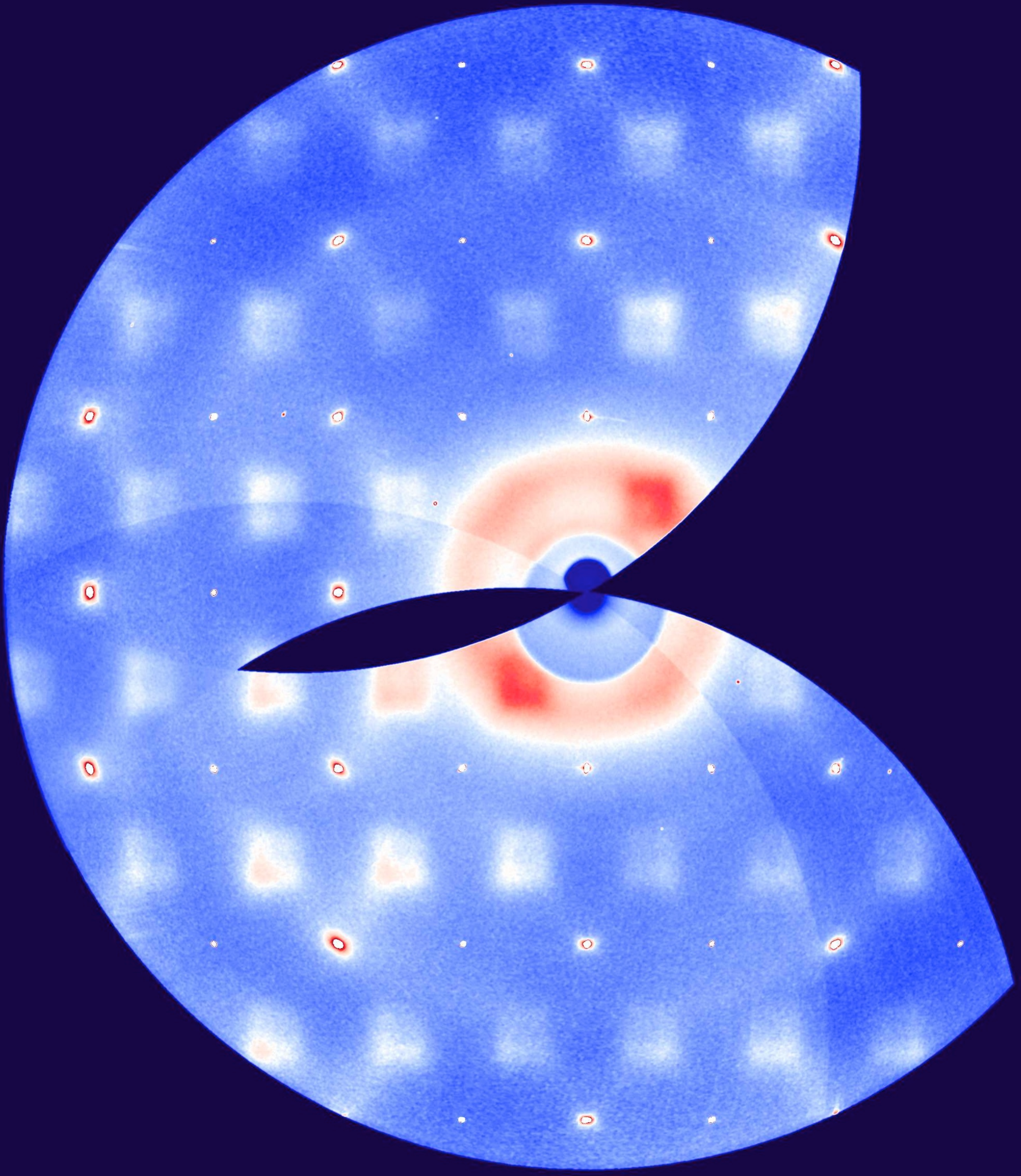}}
{\includegraphics[height=0.49\textwidth]{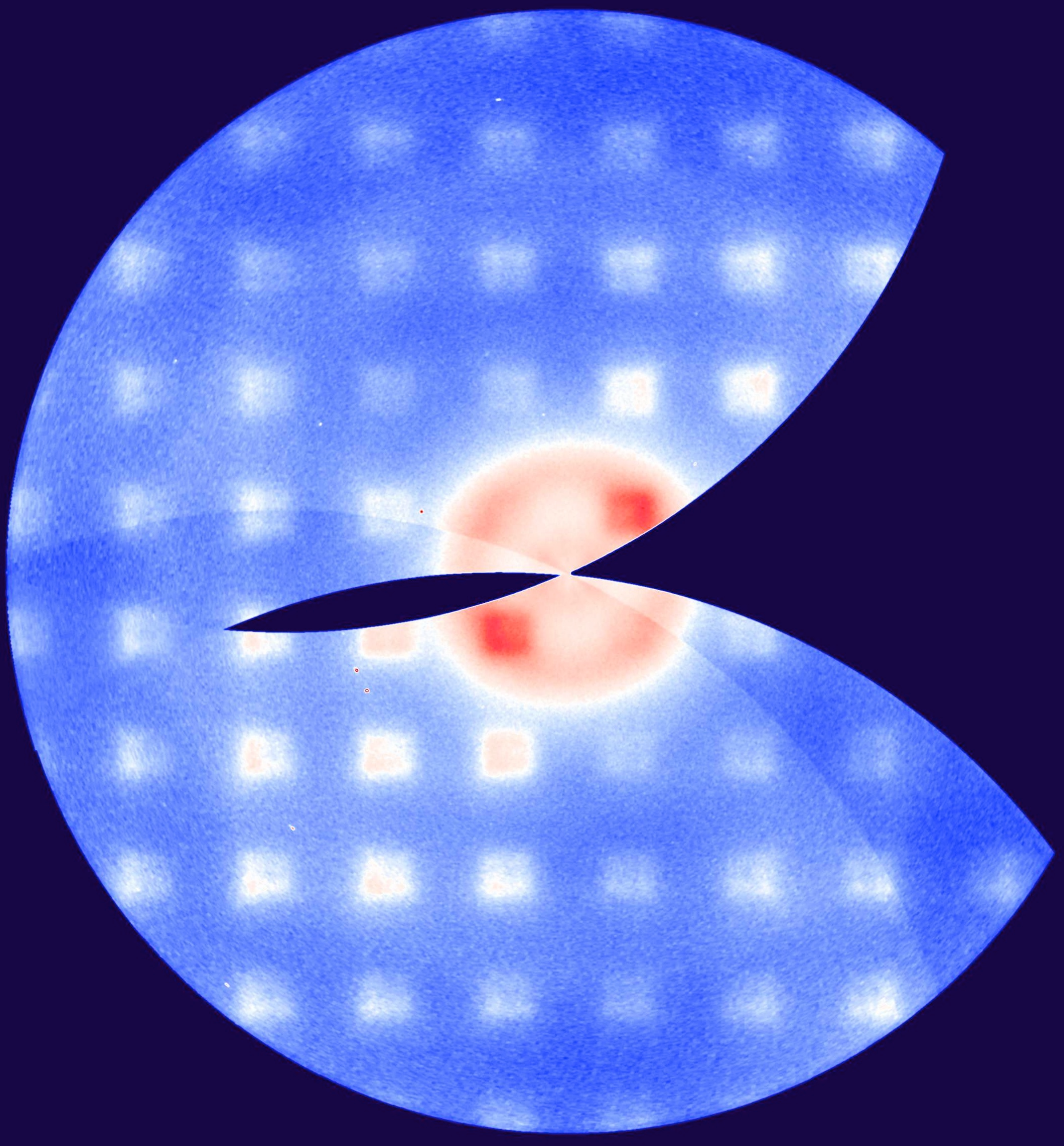}}
\end{center}
\caption{(Color online) The diffuse scattering observed for sections of reciprocal space for BMT.  
(Top left) the (100)-(001) section through the origin of the reciprocal space, showing weak diffuse streaks (here magnified) of displacive origin; (Top right) the (110)-(001) section through origin of the reciprocal space;
(Bottom) the (100)-(010) section through (0, 0, 1/2);
The square diffuse feature, that turn out to be cubes of diffuse 
scattering, has the intensity largely concentrated in the range from $(h+1/3, k+1/3,l+1/3)$ 
to $(h+2/3, k+2/3,l+2/3)$. 
}
\label{fig1}
\end{figure}

The experiments were performed on X-ray scattering instruments at the ESRF and SLS. 
At the ESRF the diffuse scattering from BMT was measured at the Swiss-Norwegian Beam Line (SNBL) and 
the XMAS beamline, and at the SLS the measurements were made at the beamline X06SA. More details of the measurements 
and the experimental results for BMT are given in Ref.~\cite{cervellino_I}. In this paper we concentrate on the 
unusual diffuse scattering observed from BMT and typical results are shown in Fig.~\ref{fig1}. Similar results were 
obtained from the SLS diffractometer X06SA, but from the XMAS data the scattering contained sharp peaks at the 
(H+1/3, K+1/3,L+1/3) positions showing the structure was trigonal. A BMT powder sample was also 
studied at the SLS on the beamline X04SA. It was revealed that the trigonal satellites were significantly broader than 
the intense peaks of the cubic phase showing that the domain size was about 20 nm for the trigonal structure. 
These results support the suggestion that our powder and the surfaces of relatively large BMT crystals consist to some extent of 
trigonal-structured domains while the bulk of crystals is cubic and shows the diffuse scattering 
of Fig.~\ref{fig1}.

This paper concentrates on the discussion and modeling of the unusual cube-shaped distribution of scattering 
and on how the different structures of BMT can exist. It is suggested that the structures are unable to relax when 
rapidly cooled from high temperature. 

BaMg$_{1/3}$Ta$_{2/3}$O$_{3}$ is cubic with lattice parameter $a=0.404$~nm. The A cation (Ba) is 
located within the unit cell in (1/2,1/2,1/2), the mixed B-site is in (0,0,0) and the three oxygens 
occupy the (0,1/2,1/2), (1/2,0,1/2), (1/2,1/2,0) positions.

\section{Idealized model of diffuse intensity}\label{hereCDF}

In this section a simple mathematical model of the diffuse intensity is described.
The model is simple because each atom is assumed to be a point scatterer , with element Xx having 
a constant scattering length $b_{\mathrm{Xx}}$. This means neglecting the atomic form factors and 
Debye-Waller factors, whose effect is just a smooth decay of the intensity at large transferred 
momentum values. 

\subsection{Idealized sharp cubes}

The scattering from BMT has diffuse scattering cubes centered at $\vqr=(h,k,l)+(\haf,\haf,\haf)$ with half-side 
1/3 (all in rlu). A simple mathematical description of their shape can be given as follows. 
In one dimension, define a box function
\begin{eqnarray}
 B_w(x) &=& 1 \quad\text{if}\ |x|<w,       \nonumber \\
        &=&   \qquad 0\ \qquad \text{otherwise}   \nonumber \\
\end{eqnarray}

Then a periodic sequence of such boxes centered on the half-integers is 
\[
\Phi_w(x)=\mathop{\sum}_{n=-\infty}^{+\infty}B_{w}(x-n-1/2)
\]
Multiplying three such terms (one for each coordinate) and adding a constant background 
(that we will show to be necessary) we obtain a model of the X-ray diffuse intensity:
\begin{eqnarray}
    I_{dif}(\vqr)&=&I_B+I_0\Phi_{1/6}(\vk{q}),\label{Icube1}\\
\text{where}&&\nonumber\\
  \Phi_{1/6}(\vk{q})&\equiv&\left\{
\begin{array}{ll}
1 &\text{within the cubes} \\
0&\text{outside} 
\end{array}
\right. =\mathop{\prod}_{\alpha=1}^3\mathop{\sum}_{h_\alpha=-\infty}^{+\infty}B_{1/6}\lrb{q_\alpha-h_\alpha-1/2}
\label{Icube2}
\end{eqnarray}
Here $w=1/6$ is the halfwidth of the observed cubes in reciprocal lattice units. 
Taking the Fourier transform of the periodic array of cubes of X-ray diffuse intensity 
$I_{dif}(\vqr)$ (\eeref{Icube1}{Icube2}), we evaluate 
the pair correlation function as
\[
G_{dif}(\vrr)
=
\int_{\mathbb{R}^3}
\DDD{3}{\vqr} 
\EE^{-2\pi \IMA \vk{r}\cdot\vqr}I_{dif}(\vqr)
\]
Note that, as $I_{dif}$ is periodic and its Fourier transform
is necessarily a weighted set of Dirac deltas on the direct lattice\footnote{This is similar in reverse to the case of a perfect crystal, where perfect periodicity in direct space means that the intensity is concentrated in the Bragg peaks, that are a weighted set of Dirac deltas located on the reciprocal lattice. The restriction of the periodic pair correlation function to one unit cell 
can be then evaluated as discrete Fourier transform of the Bragg intensities and it is named the "Patterson function".} 
\[
G_{dif}(\vrr)=\eta^2\mathop{\sum}_{\vk{M}}
S_{\vk{M}}
\delta\lrb{\vrr-\vk{M}
}
\]
where $\eta^2$ is a positive constant related to the actual scattering lengths (see \sref{sec:Bord}, \eref{etaval})  
and $S_{\vk{M}}$ are the correlation coefficients. In fact, substituting 
$I_{dif}$ from \eeref{Icube1}{Icube2}, we obtain
\begin{equation}
G_{dif}(\vrr)
=
I_B\delta(\vrr)+
\frac{I_0}{27}\mathop{\sum}_{\vk{M}}
\delta\lrb{\vrr-\vk{M}
}
\mathop{\prod}_{\alpha=1}^3\left[(-1)^{m_\alpha}
\frac{\sin\lrb{\pi m_\alpha/3}
}{\pi m_\alpha/3
}\right]
\label{eq:corr}
\end{equation}
whereas the correlation coefficients are given by
\begin{eqnarray}
S_{\vk{M}}&=&\frac{I_0}{27\eta^2}\mathop{\prod}_{\alpha=1}^3\left[(-1)^{m_\alpha}
\frac{\sin\lrb{\pi m_\alpha/3}
}{\pi m_\alpha/3
}\right],\qquad {\vk{M}}\neq(0,0,0); 
\label{eq:corr1}\\
S_{(0,0,0)}&=&
\DSF{27I_B+I_0}{27\eta^2}
\label{eq:corr2}
\end{eqnarray}

Here $\vk{M}=(m_1,m_2,m_3)$ are direct lattice nodes, where the density (and the pair correlation function)  
are different from zero. 

The distance $r=\left[m_1^2+m_2^2+m_3^2\right]^{1/2}$, 
so that $G_{dif}(r)\sim r^{-\alpha}$, with $1\leqslant \alpha\leqslant 3$, depending on the direction. 
This is a very rare case of power-law correlation decay 
of structural origin. In quite exotic systems, including spin glasses and nematic crystals, 
such correlations have been described~\cite{Feldman00,Feldman01} and named 
as quasi-long range order (QLRO). We shall name QLRCO (as quasi-long-range cation order) the kind of 
order exhibited by the BMT sample. 
Two parameters are often used to characterize the degree and extension of the structural chemical order. 
The first is the long-range order parameter of Refs.~\cite{BraggWilliams34,Kubo5}, see Appendix, 
\sref{lrop}. The value of the long-range order parameter $X = 1$ characterizes 
periodically ordered structures, as the type iii) above, while zero is the value for randomly ordered structures, 
as type i). 
Conversely, the short-range order parameters of Ref.~\cite{JMCowley34} characterize the local degrees of order - 
supposedly when long-range order is absent. We obtain (see \sref{corr:lro}, \sref{chem:sro}) that both 
long- and short-range parameters indicate a substantial degree of order, 
which is an unusual occurrence.

\subsection{Smoothed cubes}
\label{sec:smooth}

\begin{figure}[!htb]
\begin{center}
{\includegraphics[width=0.49\textwidth]{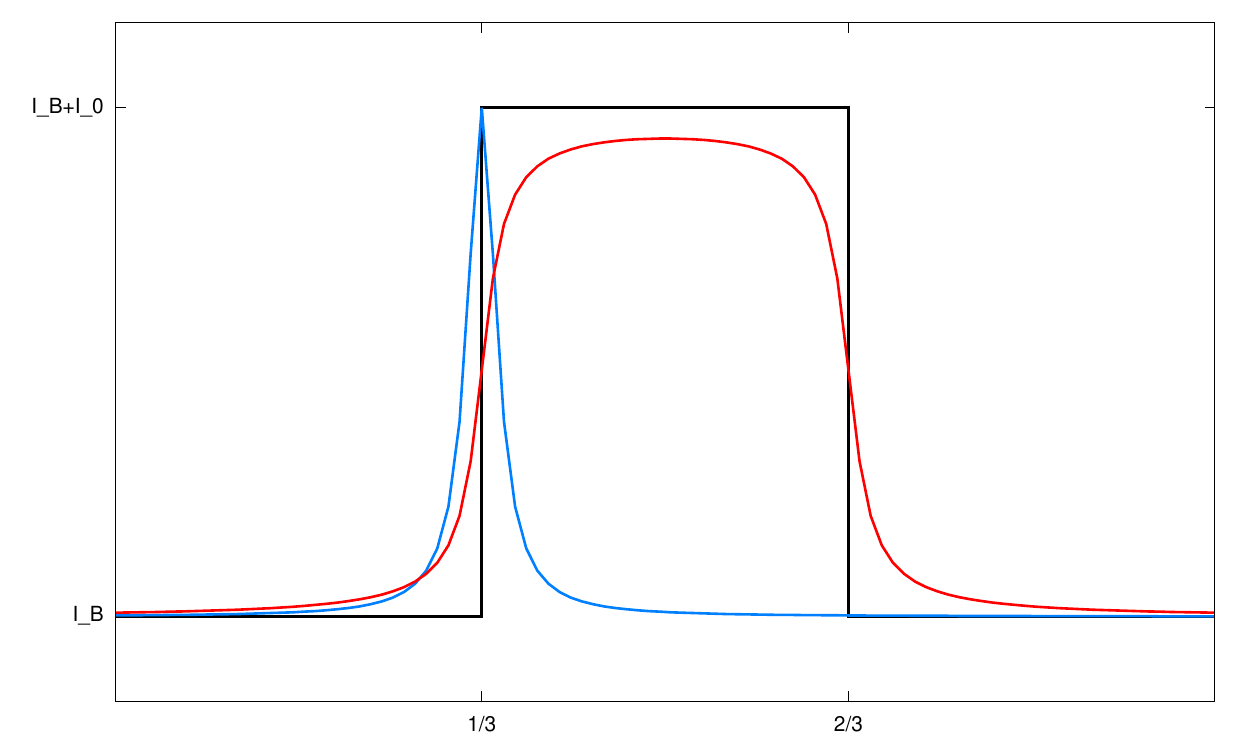}}
\end{center}
\caption{(Color online) Effect of a Lorentzian PSF on the cube shape in one dimension. Black - the sharp cube profile, 
blue - the PSF, red - the smoothed cube profile.
}
\label{fig3}
\end{figure}

The model of the diffuse intensity described above can be improved in one significant aspect. 
The cubes as observed are not perfect, as there are intensity fluctuations within them. 
Their edges, especially, are not very sharply defined. 
This  effect can be included by convoluting $\Phi(\vk{q})$ with a point-spread function (PSF) $D(\vk{q})$. 
Accordingly, the pair correlation function $G_{dif}(\vrr)$ is multiplied by the 
Fourier transform of the PSF, $\widetilde{D}(\vk{r})$ as shown in fig.~\ref{fig3}. 
The latter will be named here the \emph{correlation decay function} (CDF). 
This will give a stronger decay law for the pair correlation function. This is just 
what happens with every known diffraction phenomenon - as far as infinite crystals do not exist, 
even the most perfect standard crystal will show Bragg peaks convoluted by a suitable PSF 
and pair correlations damped by the corresponding CDF. 

We will not, in the following, burden the notation by including a PSF, 
but we will calculate some of its effects in \sref{EEsec}. 
Different possible PSF are discussed in \sref{app}. 
Here, as an example, we consider
a Lorentzian PSF along the cubic axes (Fig.~\ref{fig3})
\begin{equation}
D(\vk{q})=(\pi k_0)^{-3}\mathop{\prod}_{\alpha=1}^3\lrs{1+\lrb{q_\alpha/k_{0}}^2}^{-1}
\label{examPSF}
\end{equation}
Its Fourier transform is
\begin{equation}
\widetilde{D}(\vk{r})=\mathop{\prod}_{\alpha=1}^3\EE^{-2\pi |r_\alpha| k_0}
\label{examCDF}
\end{equation}
The correlation decay length is $L=1/(2\pi k_0)$. 
From our data, taking $k_0$ as 
the rise of the half-width of the cube edges, we can coarsely estimate that $L$ is about 10 unit 
cells. This is well below the coherent domain size, 
however it remains a considerable distance when dealing with diffuse scattering. 

As the PSF $D(\vk{q})$ is convoluted with the scattered intensity, the CDF $\widetilde{D}(\vk{r})$ 
is multiplied by the pair correlation function, that is the Fourier transform of the intensity. 
Its effect is then to limit the spatial extension of the pair correlation function. 
This is a familiar issue in crystallography. Crystal size can be estimated from the broadening of the 
Bragg peaks, due to the limited extension of the crystal - and, in turn, of the  
pair correlation function. 

We wish to remark here that the 'cube' scattering is quite different from Bragg scattering with 
respect to its response to a limitation of correlation length. In fact, convoluting the PSF with a 
Bragg peak gives the PSF itself. This is the reason for the familiar behaviour of Bragg peaks - 
namely, that their width is inversely proportional to the size of the correlated domain (a 
good estimate of this is $L$). The diffuse cubes, however, show a negligible change in width, second-order 
or more in $1/L$. This scattering form is therefore fundamentally different from Bragg scattering.

\subsection{B-site order and diffuse cubes}\label{sec:Bord}

To explain such diffuse scattering in terms of B-site cation order we split the scattering density $\rho(\vrr)$ as
\begin{equation}
\rho(\vrr)=\rho_{A}(\vrr)+\Delta\rho(\vrr),
\label{eq:flu}
\end{equation}
with $\rho_{A}$ the periodic average and $\Delta\rho$ a non-periodic zero-average fluctuation. 
If the B site is the only one that varies, all of Ba and O partial densities 
are included in $\rho_{A}$. 
As already mentioned, we simplify the picture by considering point-like 
atoms. Then $\Delta\rho(\vrr)$ is zero everywhere except on the B sites, 
and for B=Ta $\Delta\rho(\vrr)=\eta$; for B=Mg 
$\Delta\rho(\vrr)=-2\eta$. If $b_{\mathrm{Mg}}$  and 
$b_{\mathrm{Ta}}$ are the scattering lengths of Mg and Ta, 
respectively, then 
\begin{equation}
\eta=(b_{\mathrm{Ta}}-b_{\mathrm{Mg}})/3.
\label{etaval}
\end{equation}


As $\Delta\rho$ assumes only two values 
($\eta$ on Ta sites and $-2\eta$ on Mg sites) we can consider (see~\cite{Krivoglaz1}) 
the site concentration of one of the two species, say Mg, in each lattice node. Clearly this can be only 0 or 1.  We 
indicate it with $c_{\vk{M}}$, that is 1 if site $\bm{M}$ is 
occupied by Mg and 0 otherwise. Similarly we take $c'_{\vk{M}}$ as the Ta local concentration, 
that is 1 if site $\bm{M}$ 
occupied by Ta and 0 otherwise. As we have no vacancies, $c'_{\vk{M}}=1-c_{\vk{M}}$. 
Averaging over all sites, we have 
\[\langle c\rangle=1/3\quad\text{and}\quad\langle c'\rangle=2/3.\]
Furthermore, as $c$ and $c'$ can be only 0 or 1, 
\[
\langle c^2\rangle=\langle c\rangle=1/3\quad\text{and}\quad\langle c'^2\rangle=\langle c'\rangle=2/3,\] 
so the r.m.s. fluctuation of the concentration is given by
\[
\delta c=\sqrt{\langle c^2\rangle-\langle c\rangle^2}=\sqrt{2}/3.\] 
The fluctuation of the density is given by
\begin{equation}
\Delta\rho(\vrr)=\eta
\mathop{\sum}_{\vk{M}}
(-2c_{\vk{M}}+c'_{\vk{M}})\delta(\vrr-\vk{M})=\eta\mathop{\sum}_{\vk{M}}(1-3c_{\vk{M}})\delta(\vrr-\vk{M})
\label{flucden}
\end{equation}


\subsection{Correlation function}\label{sec:corr}

In order to study the diffuse scattering from BMT, 
we need to evaluate the pair correlation function of $\Delta\rho$: 
\begin{equation}
G_{dif}(\vrr)
\equiv
\DSF{1}{\Omega}
\int_{\mathbb{R}^3}\DDD{3}{\vrr'} \Delta\rho(\vrr')\Delta\rho(\vrr'+\vrr)
=
{\eta^2}\mathop{\sum}_{\vk{M}}S_{\vk{M}}\delta\lrb{\vrr-\vk{M}}
\label{eq:gdelta}
\end{equation}
Here 
\[
\Omega=Nv_c=N
\]
is the crystal volume, $N$ the number of lattice nodes within it, 
$v_c$ the unit cell volume (taken as 1 hereafter, hence the last equality). 
We shall equate this to the pair correlation function derived from the data, see \eref{eq:corr}. 
The rightmost form is 
because $\Delta\rho$ is a weighted Dirac lattice, and this property is invariant upon self-convolution. 
This is the same form as \eref{eq:corr}, so clearly now we have to calculate the 
$S_{\vk{M}}$ pair correlation coefficients for a given fluctuation density (\eref{flucden}) 
and to compare them with those obtained from the Fourier transform of the data (\eeref{eq:corr1}{eq:corr2}).

Substituting \eref{flucden} into \eref{eq:gdelta} gives 
\begin{eqnarray}
    G_{dif}(\vrr)&=&
\DSF{\eta^2}{N}
\mathop{\sum}_{\vk{M}}
\mathop{\sum}_{\vk{L}}
\lrb{1-3c_{\vk{L}}}\lrb{1-3c_{\vk{L}+\vk{M}}}
\delta\lrb{\vrr-\vk{M}}
\label{Greq}
\end{eqnarray}
where, comparing with \eref{eq:gdelta}, we obtain 
\begin{equation}
S_{\vk{M}} =
-1+\frac{9}{N}
\mathop{\sum}_{\vk{L}}
c_{\vk{L}}c_{\vk{L}+\vk{M}}\equiv -1+9\left\langle c_{\vk{L}}c_{\vk{L}+\vk{M}} \right\rangle_{\vk{L}}
\ ;\qquad S_{(000)}=-1+9\langle c^2\rangle=2.
\label{bongo}
\end{equation}

Now, the probability of finding a pair of Mg atoms separated by a lattice vector $\vk{M}$ is given by
\begin{equation}
P_{\mathrm{Mg}-\mathrm{Mg}}\lrb{\vk{M}}=\left\langle c_{\vk{L}}c_{\vk{L}+\vk{M}} \right\rangle_{\vk{L}}
=\frac{1+S_{\vk{M}}}{9}
\label{PMgMg}
\end{equation}
Similarly we can calculate the probabilities of all the other pairs (Mg-Ta/Ta-Mg and Ta-Ta) for any 
spacing vector $\vk{M}$, and the result depends only on the $S_{\vk{M}}$ correlation coefficients:
\begin{eqnarray}
P_{\mathrm{Ta}-\mathrm{Ta}}\lrb{\vk{M}}&\equiv&\left\langle (1-c_{\vk{L}})(1-c_{\vk{L}+\vk{M})} \right\rangle_{\vk{L}}=
\frac{1}{3}+\left\langle c_{\vk{L}}c_{\vk{L}+\vk{M}} \right\rangle_{\vk{L}}
=\frac{4+S_{\vk{M}}}{9}
\nonumber\\
P_{\mathrm{mix}}\lrb{\vk{M}}&\equiv&
\left\langle c_{\vk{L}}(1-c_{\vk{L}+\vk{M})} \right\rangle_{\vk{L}}
+\left\langle (1-c_{\vk{L}})c_{\vk{L}+\vk{M}} \right\rangle_{\vk{L}}
=
\frac{2}{3}-2\left\langle c_{\vk{L}}c_{\vk{L}+\vk{M}} \right\rangle_{\vk{L}}
=\frac{4-2S_{\vk{M}}}{9}
\label{POther}
\end{eqnarray}
where the subscript 'mix' refers to Mg-Ta pairs, in any order. 
Therefore, if we know the correlation coefficients we can calculate all the properties that 
depend on the pair interactions.

\subsection{Scale and background}\label{bkg}

Comparing \eref{eq:corr} with \eeref{Greq}{bongo}, considering $\vk{r}=0$, 
we obtain
\begin{equation}
S_{(000)}=2=\DSF{1}{\eta^2}
\left[
I_B+
\DSF{I_0}{27}
\right]
\qquad\Leftrightarrow\qquad
I_0={54}{\eta^2}-27I_B
\label{theo000}
\end{equation}%
We define a parameter $t$, so that if $0\leqslant t \leqslant 1$, then
\begin{equation}
I_B=2\eta^2(1-t); \qquad I_0=54\eta^2 t
\label{defit}
\end{equation}
This result describes how much of the total diffuse intensity in one Brillouin zone is a constant background 
and how much comes to form the cubes of diffuse scattering. 

As described by \eref{PMgMg}, $(1+S_{\vk{M}})/9$ is a probability, hence it must range between 0 and 1. 
In order for it to be positive for all ${\vk{M}}$,  
\begin{equation}
S_{\vk{M}}\geqslant -1.
\label{consS}
\end{equation}
For ${\vk{M}}=(1,0,0)$, this constraint - using \eref{eq:corr1} - implies
\begin{equation}
t \leqslant \DSF{\pi}{3\sqrt{3}}\approx 0.6046,
\label{cons_t}
\end{equation}
which tells us that a consistent fraction - at least about 40\% - of the diffuse intensity 
must be spread as a constant background while the rest forms the diffuse cubes. 
This is also a lower bound only, so possibly the background fraction is even larger. 

\subsection{Electrostatic energy as a function of the pair correlation function}\label{ee-cor}

In \eeref{PMgMg}{POther} we have calculated the probability of finding pairs of atoms 
of equal or different species (limited to the B sublattice) for each separation vector $\vk{M}$. 
That is all we need to calculate the electrostatic energy. 
In fact, the structure consists of the fixed part (the A sublattice and the oxygen sublattices) 
and the variable part (the B sublattice). A result of Ref.~\cite{Bellaiche98} is that 
the electrostatic energy does not depend on the interaction terms between the different sublattices; 
the fixed part then has no influence except for a constant term and one needs to calculate only 
the interactions between atoms of the B sublattice. 
Similarly to the scattering density (\eref{eq:flu}), 
we can separate the charge density $\rho_e(\vrr)$ into the the average part and fluctuating part
\begin{equation}
\rho_e(\vrr)=\rho_{e,A}(\vrr)+\Delta\rho_e(\vrr),
\label{eq:fluE}
\end{equation}
with $\rho_{e,A}$ the average component and $\Delta\rho_e$ 
a non-periodic zero-average fluctuation. Again, on Ta-occupied sites we will have 
$\Delta\rho_e=\eta_Z$ and on Mg sites, $\Delta\rho_e=-2\eta_Z$
with 
\begin{equation}
\eta_Z=(Z_{\mathrm{Ta}}-Z_{\mathrm{Mg}})/3.
\label{etavalE}
\end{equation}
If we neglect covalent bonding and assume the formal charges $Z_{\mathrm{Ta}}=+5$, $Z_{\mathrm{Mg}}=+2$, then we have 
$\eta_Z=1$. It can be shown~\cite{Bellaiche98} that only the fluctuation density 
$\Delta\rho_e$ - that is the only term that varies between the different possible phases of BMT - 
needs be included in the calculations. 
Therefore, the electrostatic energy is simply expressed - using \eeref{PMgMg}{POther} - as
\begin{equation}
E=\DSF{e^2\eta_Z^2}{8\pi\varepsilon\epsilon_0}
\mathop{\sum'}_{\vk{M}}\DSF{4P_{\mathrm{Mg}-\mathrm{Mg}}\lrb{\vk{M}}+P_{\mathrm{Ta}-\mathrm{Ta}}\lrb{\vk{M}}-2P_{\mathrm{mix}}\lrb{\vk{M}}}{\lrv{\vk{M}}}
=\DSF{e^2\eta_Z^2}{8\pi\varepsilon\epsilon_0}
\mathop{\sum'}_{\vk{M}}
\DSF{S_{\vk{M}}}{\lrv{\vk{M}}}
\label{calculE}
\end{equation}
the prime meaning that ${\vk{M}}=(0,0,0)$ is excluded from the summation. 
An additional factor of 1/2 was included in order not to count twice each bond. 

\subsection{Monte-Carlo solution}
\label{MCsol}

At this point we know that a distribution of B-site cations 
whose correlation coefficients are given by \eeref{eq:corr1}{eq:corr2} would produce a diffuse intensity 
as described by \eeref{Icube1}{Icube1}, that well matches the observed diffuse 
scattering cubes. We cannot, however, prove analytically that a distribution of two different cations 
on a cubic lattice with respective abundances of 1/3 and 2/3 exists that can satisfy \eeref{eq:corr1}{eq:corr2} 
(or alternatively, produce a diffuse intensity as in \eeref{Icube1}{Icube2}). In order to prove that 
such a distribution exists, we decided to try to fabricate one or more by a Monte-Carlo simulation. 
In this way we could also get some structural insight by analyzing the configurations so obtained. 

\begin{figure}[!hbt]
\begin{center}
{\Large{\bf{a)}}}\\[-5mm]{\includegraphics[width=0.49\textwidth]{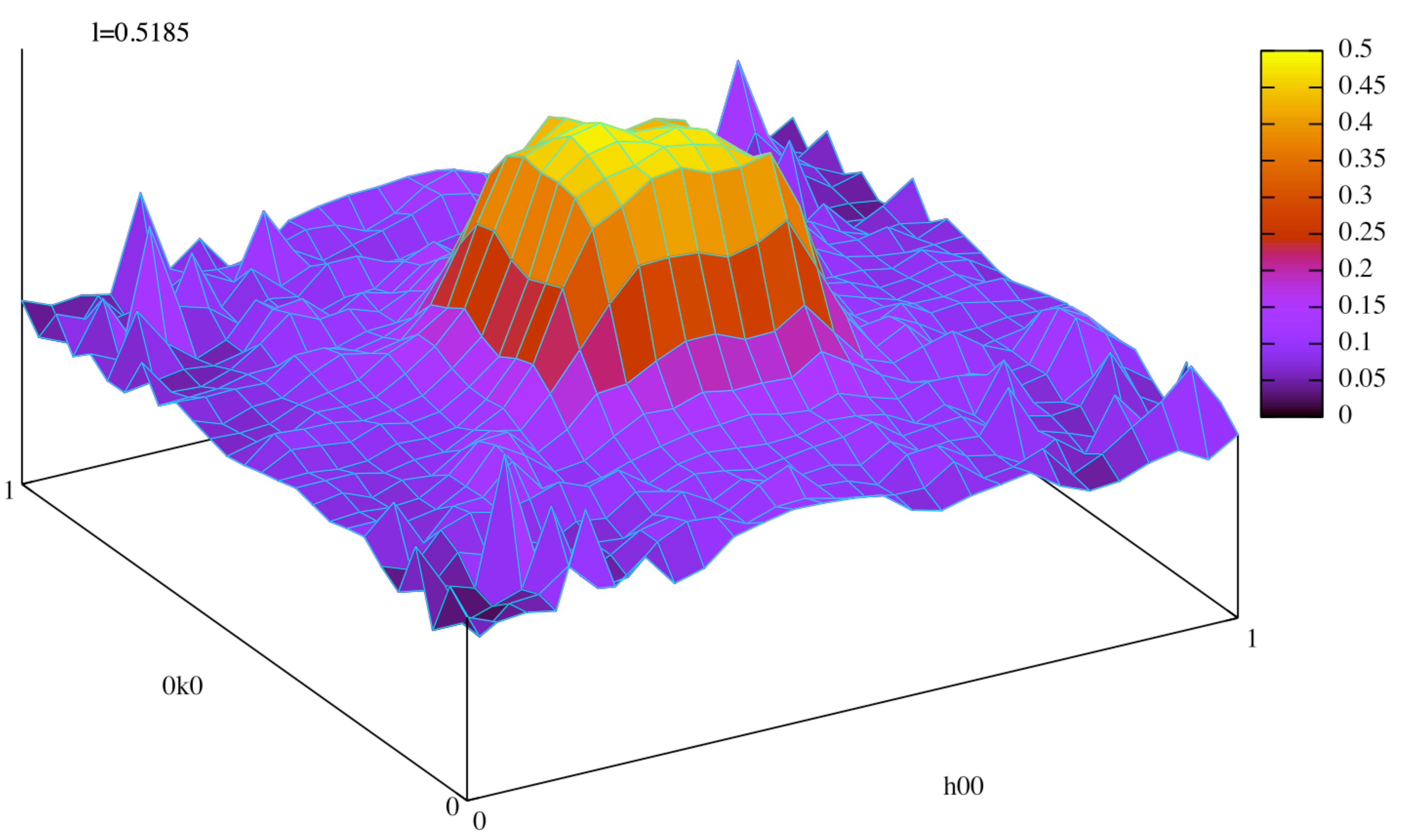}}\\
{\Large{\bf{b)}}}\\{\includegraphics[width=0.49\textwidth]{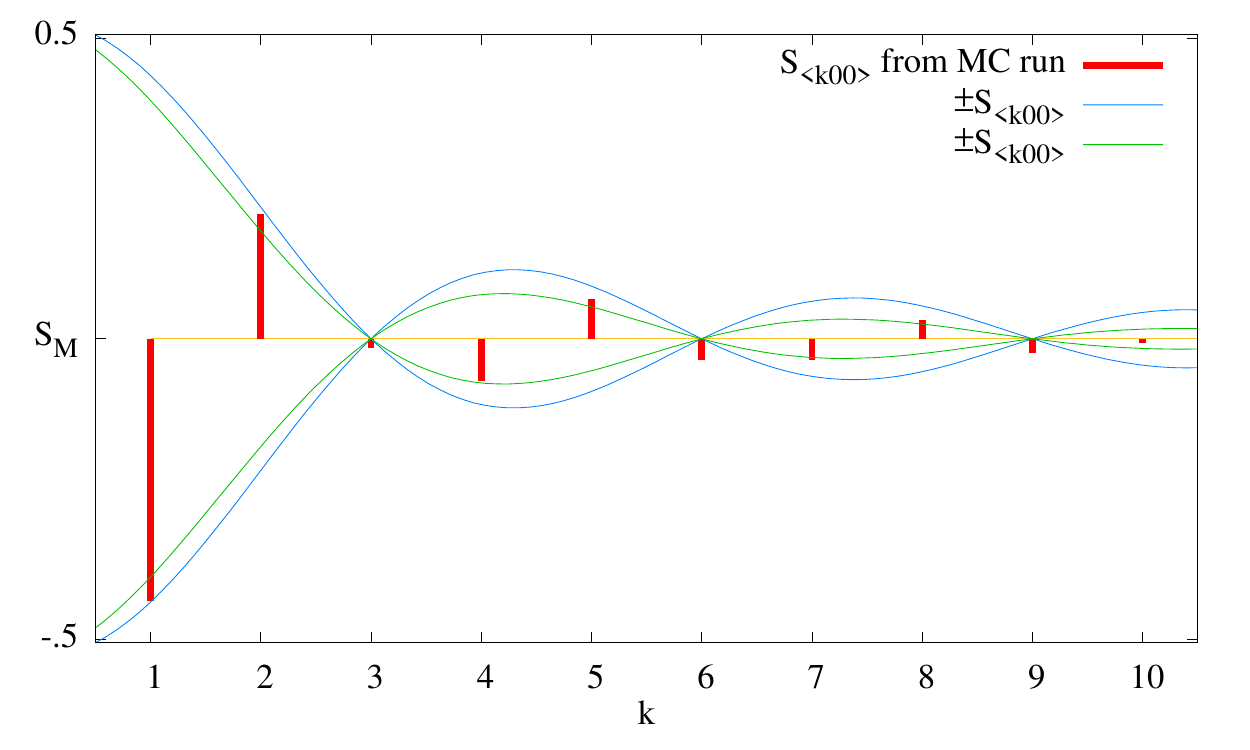}}
\end{center}
\caption{(Color online) a): The diffuse intensity calculated in one Brillouin zone from the 
Monte Carlo solution. The result shown in the $(h,k,14/27)$ plane.\\
b): Red bars: correlations $S_{\vk{M}}$ - for $\vk{M}=\langle k,0,0\rangle$ - from the 
Monte Carlo calculation. 
As allowed by the cubic symmetry, values have been averaged over $\vk{M}=(\pm k,0,0),(0,\pm k,0),(0,0,\pm k)$, and plotted \emph{vs.} $k$. 
Blue line: correlations $S_{\vk{M}}$ calculated (on the same $\vk{M}$ points) 
using 
\eeref{eq:corr1}{eq:corr2}. This is plotted as a double envelope curve (because of the factor $-1^k$) to guide the eye. 
Green line: same as the former, multiplied by a CDF $\exp(-k/L)$ as from \eref{examCDF} with $L=1/(2\pi k_0)=10$.
}
\label{qbint}
\end{figure}

A large cube ($27\times27\times27$ unit cells with the total number of atomic sites $N=19683$) 
was placed on the lattice nodes with a 'Mg' atom - represented by a scattering length $-2$ - or by a 'Ta' 
atom - represented by a scattering length $1$. These values are chosen so that the average density 
(\emph{cf.} \sref{sec:Bord}, \eref{eq:flu}) $\rho_{A}$ is zero and 
only the fluctuation density $\Delta\rho$ is not zero; 
moreover, we set $\eta=1$ for simplicity. 
The FFT power spectrum was evaluated from the intensity $I_{dif}(\vqr)$ in one Brillouin zone. 
The atoms were moved around until a good agreement with \eeref{Icube1}{Icube2} was achieved. 
The simulation was very successful. In Fig.~\ref{qbint}~a) a section of the calculated diffuse intensity 
is shown for one of the optimal configurations found. The result of the calculations compares well 
with the experimental data. (\eeref{Icube1}{Icube1}). Other sections (not plotted) are 
consistent with the one shown. 

\subsubsection{Pair correlations and long-range order parameter}\label{corr:lro}

Several equivalent configurations were constructed as above and used to evaluate the pair correlations. 
Firstly, the ratio $I_0/I_B$ for the solutions was found, and it resulted that 
$I_0/I_B\approx 3.0$. This gives a value of 
the intensity partition parameter $t\approx 0.1$ - actually much lower than its upper bound of $0.6$. 
From that, exploiting \eref{Xoft} in the Appendix, the value of the long-range order parameter 
was extracted as $X\approx 0.407$. 
We can then evaluate $I_B=1.8\eta^2$, $I_0/27=0.2\eta^2$. 

Fig.~\ref{qbint}b) shows the correlations $S_{\vk{M}}$ from the 
Monte Carlo compared with those evaluated above, for ${\vk{M}}$ along the family of 
$\langle 1,0,0\rangle$ directions). The good agreement with values calculated from \eeref{eq:corr1}{eq:corr2}, 
with and/or without an additional PSF with $L=10$, shows more quantitatively 
that the Monte Carlo converged simulation converged towards the observed pattern.

\subsubsection{Crystal chemical information and short-range order}\label{chem:sro}

\begin{figure}[hbt]
\begin{center}
{\includegraphics[width=0.49\textwidth]{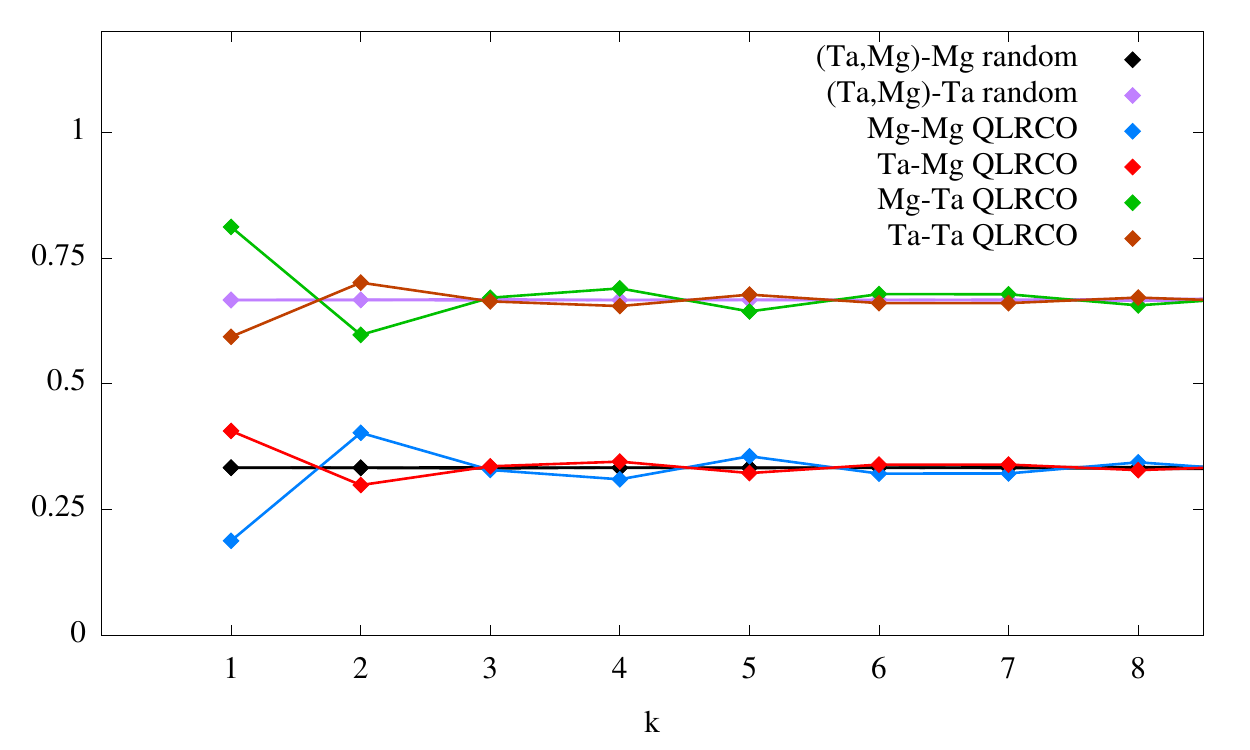}}
\end{center}
\caption{(Color online) Blue: Mg neighbor fractions for a central Mg atom in the QLRCO Monte-Carlo-generated cluster, averaged 
along all $\langle k00\rangle$ separation vectors. These correspond 
to $3P_{\mathrm{Mg}-\mathrm{Mg}}\lrb{\lra{k00}}$ (\emph{cf.} \eeref{PMgMg}{POther}, the factor 3 weighs the 
probability of the central atom being Mg). For the same cluster: 
red: Mg neighbor fractions for a central Ta atom, equal to $(3/4)P_{\mathrm{mix}}\lrb{\lra{k00}}$ 
(here factor 3/4 is obtained as product of 1/2 - because $P_{\mathrm{mix}}$ includes both bond directions, 
and 3/2 - inverse of the probability of the central atom being Ta). Green: Ta neighbor fractions for a 
central Mg atom, $(3/2)P_{\mathrm{mix}}\lrb{\lra{k00}}$. Brown: Ta neighbor fractions for a central Ta 
atom, $(3/2)P_{\mathrm{Ta}-\mathrm{Ta}}\lrb{\lra{k00}}$. 
For a cluster with randomly assigned Mg or Ta at each site, 
but with the exact composition Mg$_{1/3}$Ta$_{2/3}$, we plot in 
black the Mg neighbor concentrations of any central atom; in violet, the Ta neighbor concentrations of any 
central atom. These values can be calculated as above, simply substituting $S_{\lra{k00}}=0$.
}
\label{xxcorr}
\end{figure}

In order to display some crystal-chemical information in the MonteCarlo-obtained QLRCO configurations, 
we directly evaluated the probabilities that a Mg (respectively, Ta) atom has a Mg (respectively, Ta) neighbor at a 
distance $k$ along any of the equivalent $\lra{1,0,0}$ directions. The results are shown in Fig.~\ref{xxcorr} 
and have been calculated as follows. Each site in a cluster is treated as the central atom. 
The fraction of atoms of a given species over the neighbors having distance $k$ along the family of 
$\langle 1,0,0\rangle$ directions was calculated. Periodic boundary conditions were used. 
The fractions for all four chemical pairs 
where the atom in the origin is Mg or Ta and the neighbor is Mg or Ta, respectively. 
In Fig.~\ref{xxcorr}, we plot the 
Ta-Ta, Ta-Mg, Mg-Ta, Mg-Mg $\langle k00\rangle$ chemical pair fractions for 
the QLRCO cluster. For comparison we add the values for a cluster with randomly assigned Mg or Ta at each site, 
fulfilling a composition Mg$_{1/3}$Ta$_{2/3}$. 
These of course do not depend on the choice of the atom at the origin, and amount to 1/3 when the neighbor is 
Mg and 2/3 when it is Ta. 
From the values of the Ta-Mg (or Mg-Ta) chemical pair fractions we immediately obtain the value of the short-range 
order parameter for the first coordination shell~\cite{JMCowley34,Krayzman08}for the QLRCO phase. 
The parameter $\alpha(1)$ is found to be $\approx-0.21$. For this composition it has a maximum (in absolute value) 
at $-0.5$, for the fully ordered state, while it is zero for the random state. 
Combined with similar results obtained for the long-range order parameter 
in \sref{chem:sro} this gives a further justification for the name 'quasi-long range order'.

\section{Electrostatic energy calculation results}\label{EEsec}

In this section and the next we shall demonstrate how the electrostatic energy and the configuration entropy, 
as calculated from the structure and pair correlations for the different possible  BMT phases is sufficient 
to explain the phase diagram of BMT. In particular, the fact that the QLRCO phase is metastable at room temperature, 
and the fact that the NaCl phase, although common in similar perovskites, has never been found for BMT.

The electrostatic energy, is not the only component of the 
crystal energy, but has been previously identified as the driving force 
of the cation ordering in complex perovskites. At least it has 
been shown~\cite{Bellaiche98} that 
electrostatic energy-based predictions describe quite effectively the ordering behaviour of a large class 
of perovskites. The electrostatic energy was evaluated in a similar fashion as in Ref.~\cite{Bellaiche98}. 
As already mentioned in \sref{ee-cor}, only the fluctuation term relevant to the B-cation sublattice need to be calculated, 
because the other contributions can be considered either as fixed or as zero.

We have used a relative permittivity $\varepsilon=1$ and a cubic cell side $a=4$~\AA{} and the formal charges 
Ta$^{+5}$, Mg$^{+2}$, Ba$^{+2}$, O$^{-2}$ which gives $\eta_Z=1$. It is easy to scale the results to 
more realistic values of the parameters later on. 

The electrostatic energy was calculated using different ways. 
In all cases, the zero of the energy was fixed to the energy of a reference perovskite - 
BaTiO${}_3$ - that has a 
+4-charged cation (equal to the average charge of Mg$_{1/3}$Ta$_{2/3}$) 
on the B site. 
Electrostatic energies were calculated - apart from the reference BaTiO$_3$ perovskite, labeled uniform, for all 
the different B-site ordering discussed in this paper. The BMT phase (we label it 'QLRCO' for conciseness), 
the structure with NaCl-type partial B-site order (label 'NaCl'), 
the trigonal phase with full B-site order (label '$P{\overline{3}}m1$'), 
and the phase having fully random B-site occupancies (label 'Random') have been calculated.

As a first method of calculation, we created large clusters of atoms and computed 
the interactions directly, summing up all atom pair contributions. 
Calculations of this kind were performed for clusters of increasing size up to $27\times 27 \times 27$ cubic cells, 
at which limit the results were well stabilized and appeared to have reached their asymptotic value 
(with residual oscillations smaller than $0.1$~eV/formula unit). 
Fractional O- and Ba- site occupancies at the boundary have been used to keep the neutrality of the clusters. 
This fact~\cite{Harrison06}, together with the cubic cluster symmetry, 
explains the very rapid convergence of the electrostatic energy sum. 

As a second approach, the Madelung method~\cite{Harrison06} was used. 
This gave consistent results within $<0.1$~eV/formula unit. 

\begin{figure}[!hbt]
\begin{center}
{\includegraphics[width=0.49\textwidth]{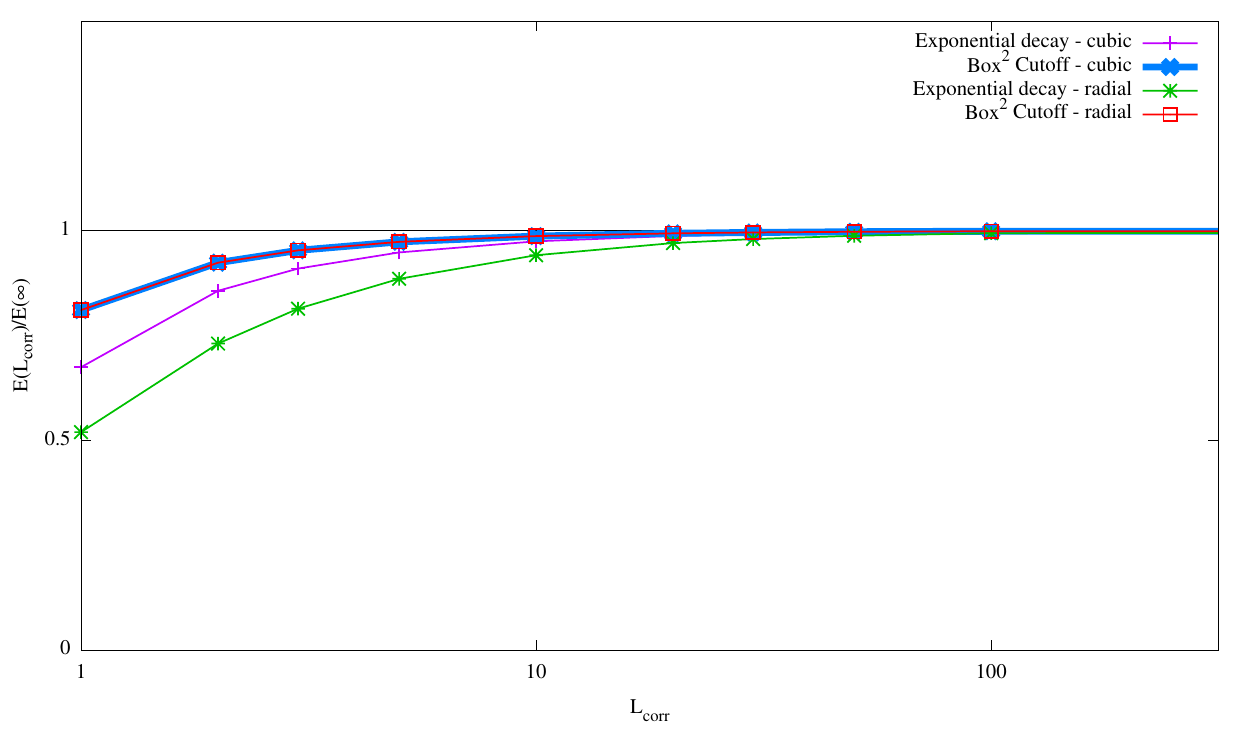}}
\end{center}
\caption{(Color online) 
Electrostatic energy - \eref{calculE} - for the QLRCO structure, when correlations are reduced by various 
decay functions (cf. \sref{hereCDF}, \sref{app}). 
}
\label{CDFfig}
\end{figure}

Thirdly, we use the knowledge of the correlation coefficients $S_{\vk{M}}$ (cf.~\sref{sec:corr}). 
In this case, the electrostatic energy is simply expressed as in \eref{calculE}. 
Such sums were evaluated in all cases from the analytical $S_{\vk{M}}$. Again, we evaluated them on 
cubes of increasing size (up to $500\times 500 \times 500$ unit cells) in order to verify the convergence - 
which was always excellent. Calculation times in this case were also extremely fast - less than one minute on a single PC 
for the largest cubes with $\approx 10^8$ nodes. 
With this latter method for the QLRCO case was also evaluated - the effect of a PSF spreading the diffuse scattering cubes 
(see Fig.~\ref{CDFfig}). It turns out that even with small correlation lengths ($L\sim 10$ unit cells) the energy was within few \%\ 
from the value as $L\rightarrow \infty$. 

For QLRCO, NaCl and Random cases (which are not deterministic) 
we have repeated the calculation for  many configurations, 
evaluating the average values and their standard deviations. 
In all cases the number of trials was chosen high enough for the standard deviations 
to be below $0.1$~eV/formula unit. 
Permittivity was taken as $\varepsilon=1$, and the 'true' energy levels can be obtained 
by dividing the results by the permittivity of the material. The results for the uniform structure 
could be compared with those evaluated using the Madelung constants 
(see~\cite{Harrison06}) and turn out to be quite accurate.  Also the electrostatic 
energy of the '$P{\overline{3}}m1$' structure - calculated by the Madelung method by 
assuming that the site geometry remains cubic is found in Ref.~\cite{Rosenstein63} 
and compares well with our results. 
Furthermore we are not interested in absolute values but in comparative ones, and these are numerically much easier 
to compute than the absolute ones. For each structure we evaluate $\Delta E_{structure}=E_{structure}-E_{\mathrm{BaTiO}_3}$. 
The results show that, with respect to the Uniform structure, in order of increasing stability:
\begin{itemize}
\item{the Random structure has the same energy within 0.1\%, 
$|\Delta E_{\mathrm{Random}}|\leqslant 0.1$~eV/formula unit;}
\item{the QLRCO structure has 1.1\% lower energy or about $\Delta E_{\mathrm{QLRCO}}=-1.902$~eV/formula unit;}
\item{the NaCl structure has 1.8\% lower energy or about $\Delta E_{\mathrm{NaCl}}=-3.229$~eV/formula unit;}
\item{the $P{\overline{3}}m1$ structure has 3.0\% lower energy or about $\Delta E_{P{\overline{3}}m1}=-5.295$~eV/formula unit.}
\end{itemize} 
Noteworthy is the level spacing ratio
\[
\zeta=\frac{\Delta E_{\mathrm{QLRCO}}-\Delta E_{\mathrm{NaCl}}
}{\Delta E_{\mathrm{NaCl}}-\Delta E_{P{\overline{3}}m1}}=0.642 <1
\]
which is smaller than
\[
\zeta'=\frac{\Delta E_{\mathrm{Random}}-\Delta E_{\mathrm{NaCl}}
}{\Delta E_{\mathrm{NaCl}}-\Delta E_{P{\overline{3}}m1}}=1.56>1
\]

There are several factors which will influence the absolute values of these energy levels, 
and the relative values of $\zeta$ and $\zeta '$ as well. These are:
\begin{itemize}
\item{The energies have to be divided by the material's relative (static) permittivity $\varepsilon$, 
that can be large in perovskites. For BaMg$_{1/3}$Ta$_{2/3}$O$_3$ it is found that $\varepsilon\approx 24$
\cite{Lin02,Lufaso04}. This factor would not change the ratio. 
}
\item{The electrostatic energy is evaluated using the full formal charges 
of the ions. Covalency corrections may play a role. 
A value of $\eta_Z=0.913$ can be extrapolated from the valence sum corrections 
evaluated for BMT in Ref.~\cite{Lufaso04}. This entails an 
absolute change of the energy levels, not affecting the ratio $\zeta$.} 
\item{The electrostatic energy is inversely proportional to the cubic lattice parameter $a$. Appropriate 
scaling is needed. However, for the known BaMg$_{1/3}$Ta$_{2/3}$O$_3$ phases, $P{\overline{3}}m1$ and QLRCO, 
$a$ (evaluated from the cubic main reflections) is the same within one part in $10^3$~\cite{Lufaso04}. 
This effect cannot change the ratios. }
\item{Geometric distortions are present in the ordered NaCl and $P{\overline{3}}m1$ structures, but they affect the distances on 
few parts in $10^4$~\cite{Lufaso04}, and as their effect on the energies is of comparable order, we can neglect them altogether.}
\item{A PSF is observed, leading to smoothed cube edges. This term is possibly 'accidental', 
depending on the detailed crystallization kinetics, so it will not be considered quantitatively in the following. 
Its effect is also small enough (see Fig.~\ref{CDFfig}) that it cannot qualitatively affect our results.}
\end{itemize}

\section{Entropy}

The crystal structure can be stabilized by the entropy. For disordered structures 
configurational entropy plays an important role. A mixed-occupancy atomic site 
with occupancy $x$ for one species and $1-x$ for another gives an entropy contribution of
\begin{equation}
S=-k_B\left[
x\log\left(x\right)+
(1-x)\log\left(1-x\right)
\right]
\end{equation}
In the $P{\overline{3}}m1$ structure no mixed sites exist, 
therefore they have no entropic contribution to the Gibbs free energy. 
There will be an entropy term related to the domain distribution, 
but that is smaller by many orders of magnitude. 

The maximal configurational entropy per formula unit is expected in the random structure and it amounts to
\begin{equation}
S_0=-k_B\left[
\frac{1}{3}\log\left(\frac{1}{3}\right)+
\frac{2}{3}\log\left(\frac{2}{3}\right)
\right]=54.85\ \mu\text{eV/K}
\end{equation}

For the QLRCO structure, as discussed, the entropy value is slightly smaller, 
but the difference can be evaluated numerically (in a simple but tedious way) 
from the correlation coefficients and the result is that the correction is negligible, so 
the entropy is $\approx S_0$.

For the NaCl structure, only 50\%{} of the sites are mixed so the entropy is $S_0/2$. 

One more important contribution is the phonon entropy. 
A different B-site degree of order can in principle affect the phonon entropy. 
However, this effect is much smaller than the configurational term. 
This is supported by the scarce literature, see \emph{e.g.}~\cite{Bogda99}. 
The specific heat $C_v$ for BMT has been measured~\cite{Gvas04}, however the phase was not fully characterized at the time 
(being, however, either QLRCO or $P{\overline{3}}m1$). We can only compare the entropy evaluated from the experimental specific heat 
with data for Pb containing relaxor perovskite (PbMg$_{1/3}$Ta$_{2/3}$O$_3$-PMT)~\cite{Gvas04,PMT2011}. The phonon entropy of PMT 
extrapolated to T$\gg\Theta_D$ exceeds that of BMT by 59 $\mu$eV/K. 
This large difference can be explained either by the polar nano-regions which are 
only present in PMT~\cite{Gvas04} or by the difference in the masses of Pb and Ba. 

\begin{figure}[!htb]
\begin{center}
{\includegraphics[width=0.49\textwidth]{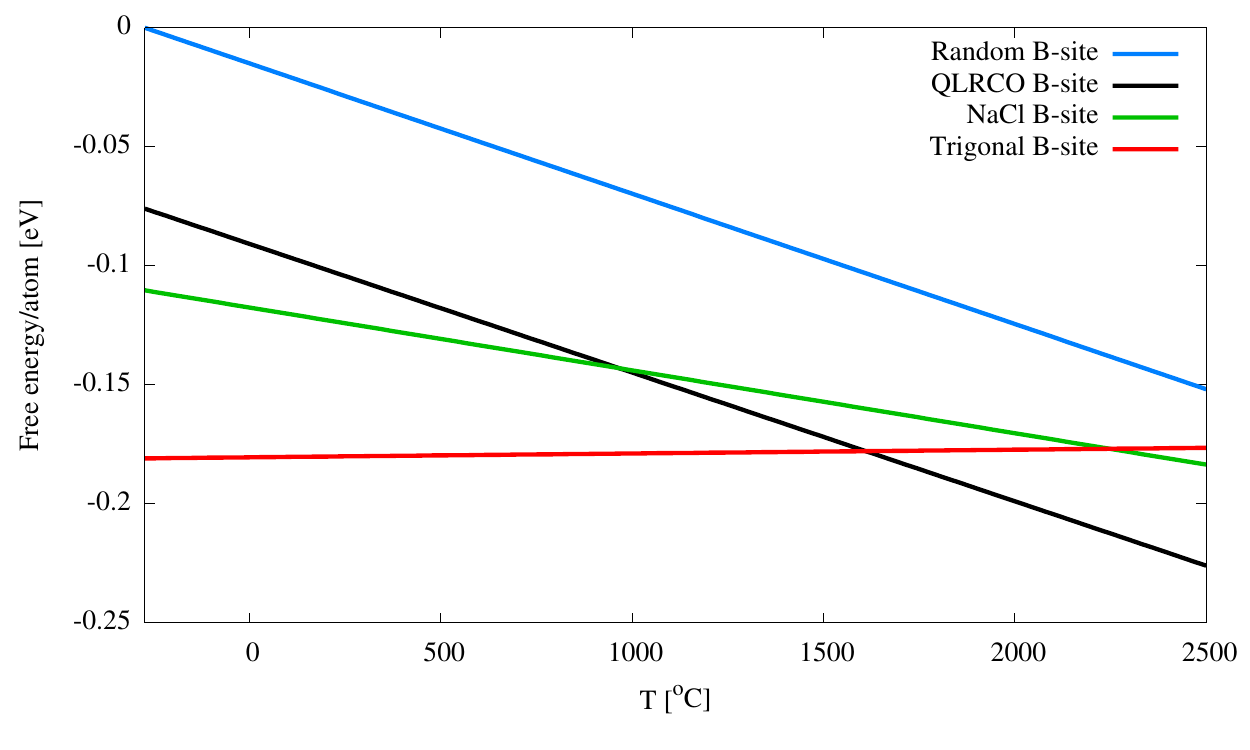}}
\end{center}
\caption{Schematic phase diagram of BMT.
}
\label{PhaseD}
\end{figure}

\section{Discussion and conclusions}

Entropy is important at high temperature for stabilizing different structures. For instance, considering the 
various free energies as calculated in the previous section, we can construct a phase diagram of BMT~(Fig.~\ref{PhaseD}). 
We have taken, as previously stated, realistic values of  $\varepsilon=24$ (assumed constant with $T$), $\eta_Z=0.913$, 
and a relative linear expansion coefficient of 9$\times 10^{-6}$. 
We have not included any contribution from imperfect ordering (namely, the effect of 
a CDF limiting the spatial extent of correlations) because it is not easy to evaluate quantitatively and it is not 
an intrinsic effect but rather a kinetic effect while its contribution is small 
(Fig.~\ref{CDFfig}) and the leading approximation would only compress the energy range.

The entropy term is canceled by the energy difference between QLRCO and $P{\overline{3}}m1$ phases at about 1600~\degC{}C, that is 
about the melting point. Above this temperature the QLRCO state would be stable while the $P{\overline{3}}m1$ state would be stable below. 
We can also see from~Fig.~\ref{PhaseD} that there is no region that is stable 
for the NaCl-ordered phase, and in fact, this phase has never been observed in BMT, at variance with PMT~\cite{Vieh95,Yan98}. 

Due to the slow kinetics associated with cation ordering, chemical ordering is usually achieved 
only after annealing for a long time at very high $T$, where the equilibrium state may be dictated by the entropy. 
Cooling - due to the slow kinetics - cannot change the high-$T$ equilibrium state 
so that it would remain preserved (metastable) at room temperature.

The random structure is always less stable than the QLRCO structure, because of the tiny 
entropy difference and the large energy difference that favours the latter structure. 
Therefore the random structure is a disordered metastable state. 
The ground state - or the thermodynamic equilibrium state at 0~K - is the $P{\overline{3}}m1$ structure. 
The true high-temperature thermodynamic equilibrium state is QLRCO, because the  
electrostatic energy difference with the $P{\overline{3}}m1$ structure is overcome by the entropy 
term at high $T$, even if the dielectric constant is of moderate size. The 
NaCl structure is intermediate between these two. Given that its Gibbs free energy decreases 
with increasing temperature half as fast as the free energy of the QLRCO state, the ratio $\zeta$ becomes crucial 
to determine if the NaCl order 
can be the thermodynamic equilibrium state in some temperature range. In 
fact for $\zeta<1$ the Gibbs free energy of 
QLRCO is always lower than that of NaCl structure, while for $\zeta>1$ there exists a temperature interval 
where NaCl structure is the equilibrium state. The QLRCO structure is favoured above that temperature and the 
$P{\overline{3}}m1$ structure below.

\section{Acknowledgments}
 We are grateful for the opportunity to carry out the measurements using the spectrometers: SNBL and XMAS at 
the ESRF, and the beamlines X04SA (MS-Powder) and X06SA at the SLS. We thank Dr. Laurence Bouchenoire for 
invaluable assistance in the XMAS experiment.


%
\section{Appendix}

\subsection{More choices of point-spread functions}\label{app}

A particular type of PSF/CDF pair has been shown in \eeref{examPSF}{examCDF}. In this section some different 
PSF-CDF pairs will be discussed. PSF's are normalized so that their sum is 1.

Firstly, as in \eeref{examPSF}{examCDF}, we consider the symmetry of the PSF to be cubic. 
A Lorentzian PSF and the corresponding exponential CDF are presented in \eeref{examPSF}{examCDF}. 
A simpler PSF would be a sinc function, corresponding to a box function CDF. 
The negative parts of the sinc tails are, however, a problem. Therefore we consider a sinc squared function:
\begin{equation}
\label{psf1}
\text{PSF:}\quad D(\vk{q})=(\pi k_0)^{-3}\mathop{\prod}_{\alpha=1}^3\DSF{\sin^2\lrb{q_\alpha/k_0}}{(q_\alpha/k_0)^2};
\qquad\text{CDF:}\quad \widetilde{D}(\vk{r})=\mathop{\prod}_{\alpha=1}^3 (1-\pi |r_\alpha| k_0) \mathrm{box}\lrb{\pi |r_\alpha| k_0}
\end{equation}


Correspondingly, if the symmetry is spherical, the  PSF/CDF are given by the radial Lorentzian or exponential:
\begin{equation}
\label{psf2}
\text{PSF:}\quad D(\vk{q})=\pi^{-2}k_0^{-3}\lrs{1+\lrb{q/k_{0}}^2}^{-2};
\qquad\text{CDF:}\quad \widetilde{D}(\vk{r})=\EE^{-2\pi r k_0},
\end{equation}
and the analog of the squared sinc/box pair:
\begin{equation}
\label{psf3}
\text{PSF:}\ D(\vk{q})=3(2\pi^2k_0^3)^{-1}\lrs{\DSF{\sin\lrb{q/k_0}-\lrb{q/k_0}\cos\lrb{q/k_0}}{\lrb{q/k_0}^3}}^2;
\quad\text{CDF:}\ \widetilde{D}(\vk{r})=(1+\pi r k_0/2)(1-\pi r k_0)^2\mathrm{box}\lrb{\pi r k_0}
\end{equation}

\subsection{Correlation coefficients for ordered and semiordered BMT phases}\label{pha}

In the case that the B-site is fully disordered, 
\emph{i.e.} when all $c_{\vk{L}}$ are randomly assigned $S_{(000)}=2$, but 
\[
S_{\vk{M}} =-1+{9}\langle c\rangle^2=0 \qquad \text{whenever}\ {\vk{M}}\neq(0,0,0)
\]
so the Fourier transform $I(\vqr)$ is a constant, $I(\vk{q})=2\eta^2 $. 

The trigonal superstructure has as the 3-fold axis one of the equivalent $\lra{1,1,1}$ cubic axes. 
This means that it exists in four different orientations, where the 3-fold axis $\bm{T}$ is $\bm{T}=(1,1,1)$, 
$\bm{T}=(1,-1,-1)$, $\bm{T}=(-1,1,-1)$, $\bm{T}=(-1,-1,1)$, respectively. The B-site superstructure is obtained by 
positioning in a lattice node $\bm{M}$ on a Mg atom if $\bm{M}\cdot\bm{T}=0\ \text{modulo}\ 3$. Therefore, we will have 
pure Mg planes alternating with pairs of Ta planes, stacked orthogonally to $\bm{T}$. 
In any crystallite, domains of the four different orientations will be present in equal volumes, therefore 
conserving the global cubic symmetry, in agreement with Landau's theory. 
The correlations can be calculated for each orientation and then averaged. 
However, one can also calculate them from the scattering pattern, as an average over domains. 
The superstructure satellites - considering equipopulated domains - are eight equal Bragg peaks 
(each with intensity $I_{P{\overline{3}}m1}$) at $(\pm 1/3,\pm 1/3,\pm 1/3)$. Then
\[
S_{\vk{M}}=(-1)^{k+1}2^k\DSF{I_{P{\overline{3}}m1}}{\eta^2}
\]
where $k$ is the number of components of ${\vk{M}}$ that are equal to zero, or 
$k=\#\{m_\alpha=0\ \text{modulo}\ 3,\quad \alpha=1,2,3\}$. 
From the condition $S_{(000)}=2$, we obtain 
$I_{P{\overline{3}}m1}=\DSF{\eta^2}{4}$, so that 
\[
S_{\vk{M}}=(-1)^{k+1}2^{k-2}
\]
The hypothetical NaCl-type superstructure is simpler. There is only one Bragg peak at (1/2,1/2,1/2) 
with intensity $I_{\mathrm{NaCl}}$  
plus a small constant background $I_b$ due to residual randomness on half of the sites. 
So one obtains
\[
S_{\vk{M}}=(-1)^{m_1+m_2+m_3}\DSF{I_{\mathrm{NaCl}}}{\eta^2}; \qquad S_{(000)}=\DSF{I_{\mathrm{NaCl}}+I_b}{\eta^2}
\]
As half the sites contribute to the background and the occupancy of the random sites is always divided as 1/3 - 2/3, 
$I_b=\eta^2 $ and again using the condition $S_{(000)}=2$, we have
\[
I_{\mathrm{NaCl}}=\eta^2.
\]

\subsection{Long-range order parameter}\label{lrop}

In this section we show how the former results can be related to the formalism of the long-range order parameter. 
A simple way of describing systems where there is a fully ordered state 
is by the long-range order parameter~\cite{BraggWilliams34,Kubo5}. 
Here we will not go to all the detail as it is a well-known theory. For BMT the ordered state 
is the trigonal modification. Here the cubic B-site lattice is divided into 3 sublattices, denoted (a),(b),(c) for convenience   
of which one is occupied by Mg and two by Ta. There are four ways of choosing the three-fold axis of the 
trigonal structure and for each three permutations of the Mg sublattice between (a),(b),(c). Suppose that 
we chose one axis and that the Mg sublattice is (a). 
In the ordered state all Mg atoms are actually on the (a) sublattice; if we have $N$ total atoms, we will have $N/3$ Mg atoms in the (a) 
sublattice. In the random state, the Mg concentration on any site will be the same, so we will find only $N/9$ Mg atoms in the (a) 
sublattice. We can parametrize this situation if we say that the number of Mg atoms on the (a) sublattice is given by
\(
{N}(1+2X)/9
\)
where $X$ is a parameter that is 1 in the ordered state, 0 in the random one and assumes intermediate values for all intermediate 
states. 
Using the same $X$, we can parametrize the number of Ta atoms on the (b)+(c) double sublattice, and all cross-frequencies. 
More involved calculations (see~\cite{Kubo5} for details of the method) lead to the expressions for the nearest-neighbor pairs frequencies 
as a function of $X$. If the interaction is fundamentally short-ranged - or if the interactions beyond the first neighbors are not 
very important - one can then use different interaction hypotheses and calculate the internal energy. 
Finally, from the stochastic occupancy of each site, it is possible to evaluate the configurational entropy. 
The nearest-neighbor bond fractions 
turn out to be 
\[
\nu_1=\frac{1}{9}(1-X^2)\qquad \text{Mg-Mg pairs}
\]\[
\nu_2=\frac{1}{9}(4-X^2)\qquad \text{Ta-Ta pairs}
\]\[
\nu_3=\frac{2}{9}(2+X^2)\qquad \text{Ta-Mg pairs}
\]

As nearest-neighbor bond vectors are just the $\langle1,0,0\rangle$ family, 
it is easy to see that 
\[
\nu_1=\frac{1}{N}\mathop{\sum}_{\vk{L}}
c_{\vk{L}}c_{\vk{L}+\langle1,0,0\rangle} = \DSF{1+S_{\langle1,0,0\rangle}}{9}\ .
\]
Similar relations for the other $\nu$'s can be given using the $c'_{\vk{L}}=1-c_{\vk{L}}$'s. 
Since $X^2=-S_{\langle1,0,0\rangle}$, and using \eeref{eq:corr1}{defit}, a relationship is found 
between $X$ and the background parameter $t$ (see \sref{bkg}) 
\begin{equation}
X=\lrs{
\DSF{
3t\sqrt{3}
}{
\pi
}
}^{1/2}
\label{Xoft}
\end{equation}
The correlation coefficients are evaluated as in \sref{pha}, while
$X=1$ for the trigonal phase but also for the partially ordered hypothetical NaCl-type phase. 
Obviously $X=0$ for the random B-site phase, while for the QLRCO phase, 
using the $t\approx 0.1$ value estimated from the MonteCarlo runs, we have $X=0.407$, which is well above 
the value of zero for the completely random structure. 
This - combined with similar results obtained for the short-range order parameter of~\cite{JMCowley34} 
in \sref{chem:sro} - is a further justification for the name 'quasi-long range order'. 


\end{document}